\documentclass[12pt]{article}

\usepackage{scicite}
\usepackage{colortbl}
\usepackage{times}

\usepackage{graphicx}%
\usepackage{multirow}%
\usepackage{amsmath,amssymb,amsfonts}%
\usepackage{xcolor}%
\usepackage{tikz}
\usepackage{scalerel}
\usepackage{colortbl}
\usepackage{hyperref}
\usepackage[normalem]{ulem} %INCLUDE BY ERICK: TO STRIKETHROUGH

\usepackage[percent]{overpic}

\usepackage{xcolor}
\usepackage{graphicx}%

\newenvironment{sciabstract}{%
\begin{quote} \bf}
{\end{quote}}

\title{Resultant force on grains of a real sand dune: How to measure it?} 
\author
{Renato F. Miotto,$^{1\dagger}$ Carlos A. Alvarez,$^{2}$ Danilo S. Borges,$^{1}$\\
 William R. Wolf,$^{1}$ Erick M. Franklin$^{1}$\\
\\
\normalsize{$^{1}$Faculdade de Engenharia Mec\^anica, Universidade Estadual de Campinas (UNICAMP),}\\
\normalsize{Rua Mendeleyev, 200, 13083-860, Campinas, Brazil}\\
\normalsize{$^{2}$Department of Earth and Planetary Sciences, Stanford University}\\
\normalsize{450 Jane Stanford Way, Bldg. 320, Stanford, 94305, CA, USA}
\\
\normalsize{$^\dagger$To whom correspondence should be addressed; E-mail:  miotto@fem.unicamp.br}
}

\date{}

\begin{document} 


\baselineskip24pt


\maketitle


\begin{sciabstract}
  % Dunes are bedforms that appear on sandy terrains sheared by a fluid flow, being found on Earth (both in eolian and aquatic environments), Mars, and other celestial bodies; however, although ubiquitous in nature, their dynamics is not fully understood. One major problem is acquiring data at the grain scale, given the large number of grains forming a dune: quadrillions for a terrestrial desert dune, and even more for Martian dunes. Here, based on subaqueous experiments using a high-speed camera, discrete numerical computations solving the motion at the grain scale, and a special training of a convolutional neural network, we show that it is, in fact, possible to estimate the resultant force acting on the grains of a dune by using images. This procedure is a breakthrough that opens new possibilities for measuring the resultant force not only on the grains of a dune, but also on relatively small elements that are imaged over time, such as rocks, boulders, rovers, and human-built constructions photographed by satellites on terrestrial and Martian landscapes.
  %
  Dunes are bedforms found on sandy terrains shaped by fluid flow on Earth, Mars, and other celestial bodies. Despite their prevalence, understanding dune dynamics at the grain scale is challenging due to the vast number of grains involved. In this study, we demonstrate a novel approach to estimate the forces acting on individual dune grains using images. By combining subaqueous experiments, high-speed camera recordings, discrete numerical simulations, and a specially trained convolutional neural network, we can quantify these forces with high accuracy. This method represents a breakthrough in studying granular dynamics, offering a new way to measure forces not only on dune grains but also on smaller objects, such as rocks, boulders, rovers, and man-made structures, observed in satellite images of both Earth and Mars. This technique expands our ability to analyze and understand fluid-grain interactions in diverse environments.
\end{sciabstract}

% In setting up this template for *Science* papers, we've used both
% the \section* command and the \paragraph* command for topical
% divisions.  Which you use will of course depend on the type of paper
% you're writing.  Review Articles tend to have displayed headings, for
% which \section* is more appropriate; Research Articles, when they have
% formal topical divisions at all, tend to signal them with bold text
% that runs into the paragraph, for which \paragraph* is the right
% choice.  Either way, use the asterisk (*) modifier, as shown, to

\section*{Introduction}

Sand dunes are formed by the excavation and deposition of sand in troughs and on crests, respectively, due to the action of a fluid flow \cite{Bagnold_1}, being found on Earth (both in eolian and aquatic environments), Mars, and other celestial bodies \cite{Herrmann_Sauermann,Hersen_3,Elbelrhiti,Claudin_Andreotti,Parteli2,Courrech}. However, the detailed mechanisms for excavation and deposition are intricate, depending on local variations of sand motion. These, in their turn, depend on fluid flow disturbances and the flow regime, relaxation and inertial mechanisms of granular motion, and grain granulometry, to name but a few parameters \cite{Franklin_12}. In particular, the fact that eolian and Martian dunes consist of large collections of discrete particles (quadrillions for each eolian dune, and even more for Martian dunes) makes it difficult to measure their dynamics at the grain scale. Therefore, although ubiquitous in nature, the dynamics of sand dunes is far from being fully understood.

\begin{figure}[!htb]
    \centering
    \begin{overpic}[width=0.325\textwidth]{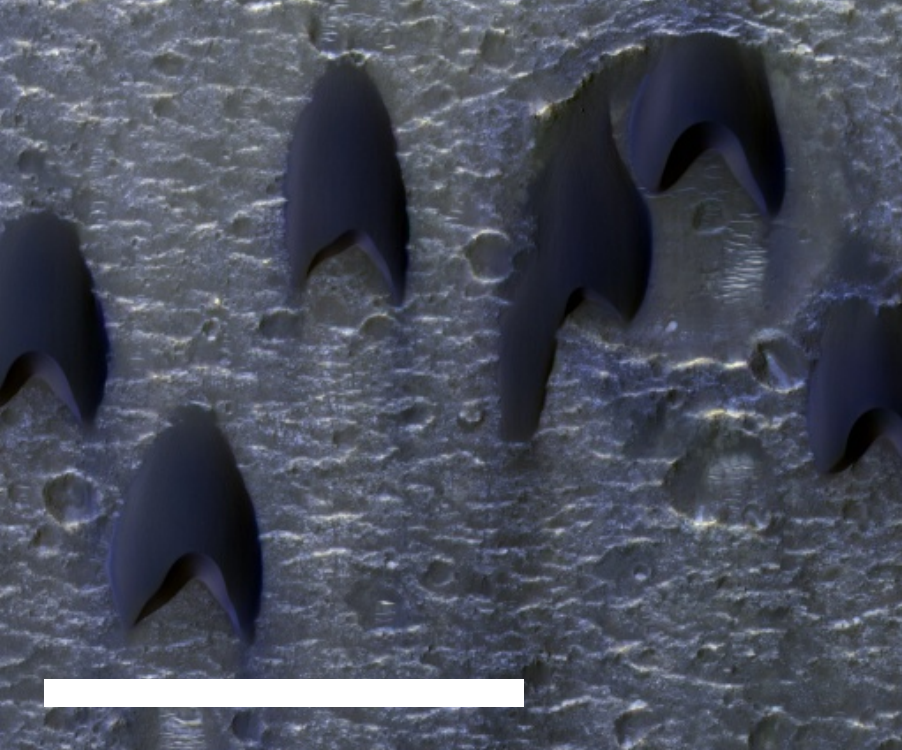}
        \put(3, 73){\textbf{\textcolor{white}{(a)}}}
        \put(10, 10){\textcolor{white}{\scriptsize 100 m}}
    \end{overpic}
    \begin{overpic}[width=0.325\textwidth]{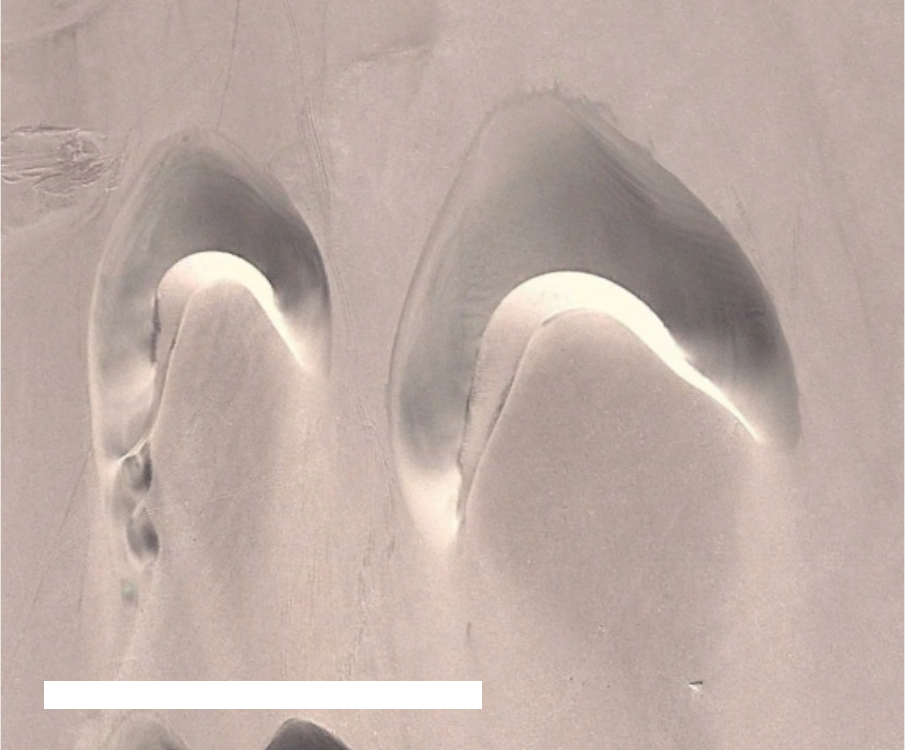}
        \put(3, 73){\textbf{\textcolor{white}{(b)}}}
        \put(10, 10){\textcolor{white}{\scriptsize 200 m}}
    \end{overpic}    
    \begin{overpic}[width=0.325\textwidth]{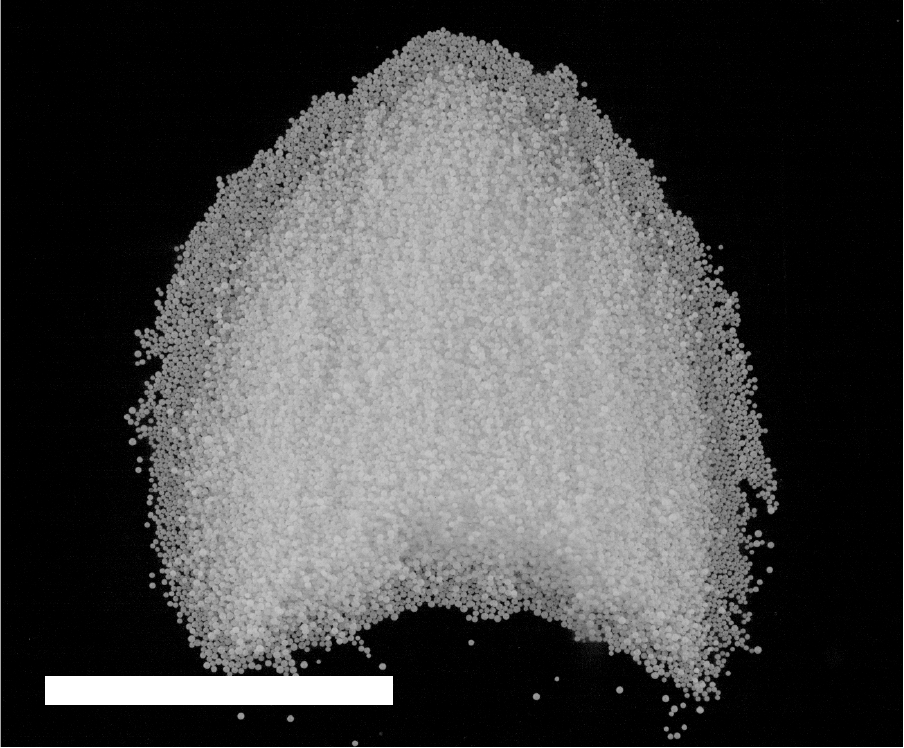}
        \put(3, 73){\textbf{\textcolor{white}{(c)}}}
        \put(10, 10){\textcolor{white}{\scriptsize 3 cm}}
    \end{overpic}   
    % https://www.uahirise.org/ESP_034815_2035
    \caption{(a) HiRISE image \cite{Supplemental_HIRISE} showing a field of barchans undergoing complex interactions on the surface of Mars: 23.190$^\circ$ latitude (centered), 339.585$^\circ$ longitude (East), spacecraft altitude 287.3 km. Courtesy NASA/JPL-Caltech/UArizona; (b) Field of barchans on the Nazca desert: -15.278$^\circ$ latitude, -74.878$^\circ$ longitude, June 2012. Courtesy Google Earth Pro; (c) Experimental image of barchan obtained in water channel.}
    \label{fig:satelite}
\end{figure}

Among different types, crescent-shape dunes with horns pointing downstream, known as barchans, are found under roughly one-direction flows and when the amount of available sand is limited \cite{Bagnold_1}. The crescent shape of barchans is a strong attractor, being approximately the same whether in aquatic or eolian environments, whether on Earth or Mars (as seen in Figs. \ref{fig:satelite}a and \ref{fig:satelite}b, for example), or in a laboratory experiment performed in a water channel (Fig. \ref{fig:satelite}c). The scales, however, are highly dependent on the environment barchans are exposed to, in particular the state of the fluid: centimeters and minutes for subaqueous barchans, hundreds of meters and years for eolian barchans on Earth, and up to kilometers and millenniums for Martian barchans \cite{Claudin_Andreotti,Hersen_1}. Besides being frequently found in nature, barchans are of special interest for their horns indicate the orientation of the mean flow. This is useful for deducing from satellite images the average wind direction over the last decade, century, or even thousands of years on terrestrial and Martian fields \cite{Rubanenko,Rubanenko2}, and estimating how global changes are affecting Earth, based on the dynamics of dunes \cite{Baas}.

Several experimental studies have been carried out in water tanks and channels, taking advantage of the smaller and faster scales of subaqueous barchans, which allowed the investigation of the morphodynamics of dunes when isolated \cite{Hersen_1,Endo,Hori,Alvarez3,Alvarez4,Yang_3} or interacting with each other \cite{Assis,Assis2,He2}. In particular, Refs. \cite{Alvarez3,Alvarez4,Assis2} performed experiments at the grain scale, in which they measured trajectories of individual grains and computed their fluxes. Based on the same scaling principle, numerical simulations at the grain scale were conducted for subaqueous barchans \cite{Alvarez5,Alvarez7,Lima2}. In these simulations, in addition to grain trajectories, the authors measured the forces acting on each individual grain, from which they computed statistics and showed the distribution of average forces within a barchan dune. The latter information has never been measured in experiments, even for the smaller ($\approx$ 100.000 grains) subaqueous barchans, for which one would require a large number of tiny accelerometers, or develop a new technique. For eolian dunes, the number of accelerometers would be even more prohibitive.

On the other hand, barchans on Earth and Mars have been monitored over the last decades by using satellite images \cite{Silvestro,Fenton}. For that, automatic detection has been crucial given the large number of barchans on both planets. The first works used manual detection \cite{Tsoar,Tsoar2,Zhang_2} (codes specially written for detection and classification), while more recent studies made use of machine and deep learning (ML and DL, respectively) for automatic detection \cite{Azzaoui,Carrera,Rubanenko,Rubanenko2}. Recently, Rubanenko et al. \cite{Rubanenko} trained a Mask R-CNN (regional convolutional neural network) \cite{He} for detecting, classifying, and outlining barchan dunes on Mars and Earth, and, with that, mapped a great part of the Martian surface and detected isolated barchans (showing, for example, that around 30 and 60\%  of dune fields on the southern and northern hemispheres, respectively, are covered with barchans). Later, Rubanenko et al. \cite{Rubanenko2} used the same technique for showing that m-scale ripples found on Mars vary with the fluid density, being caused by hydrodynamic instability instead of surface creep (the secondary motion due to the impact of saltating grains, also known as reptation). More recently, C\'u\~nez and Franklin \cite{Cunez4} successfully trained a CNN for identifying, classifying, outlining, and tracking barchans along different images, for different environments and image types, even when they underwent complex barchan-barchan interactions. Although satellite imagery increased considerably our understanding of the Martian and terrestrial landscapes, measurements were limited to the bedform scale (lengths, widths, areas). At least until now.

In this paper, we introduce a new concept for measuring inaccessible quantities by using deep learning. For that, we carried out experiments where barchans in subaqueous environment were filmed with a high-speed camera, and discrete numerical computations that solved the motion of each grain with, as one of outputs, images of grains colored in accordance with their resultant force. Afterward, we trained a convolutional neural network (CNN) with the simulation outputs with and without colors (the latter, which correspond to simple top-view images of dunes, are the inputs of the CNN, and the former the outputs), and the CNN was able to identify the colors (forces) on outputs of new simulations with good accuracy. Therefore, the CNN can learn the force distribution along the dunes based on their morphological features, being able to extrapolate solutions for dunes never seen by the network. Finally, when applied to images from experiments at the same conditions of our simulations, the CNN estimated the forces on groups of grains. As one day the digital elevation model %(DEM)
\cite{Li,Okolie} was proposed as a new tool for measurements from aerial images, we now propose a new method for measuring small scale quantities based on satellite and aerial images. This method is a breakthrough that opens new possibilities for measuring not only the resultant force acting on the grains of a dune, but also different quantities related to small elements on a given surface, such as grains, rocks, rovers and human-built constructions on terrestrial and Martian landscapes photographed by satellites.% The outcome of this research can have considerable impact on agriculture\cite{Baas}, housing \cite{WoodTV}, forest conservation \cite{Baas}, and planetary exploration, to cite but a few examples.

\begin{figure}[!ht]
	\centering
	\includegraphics[width=0.99\textwidth]{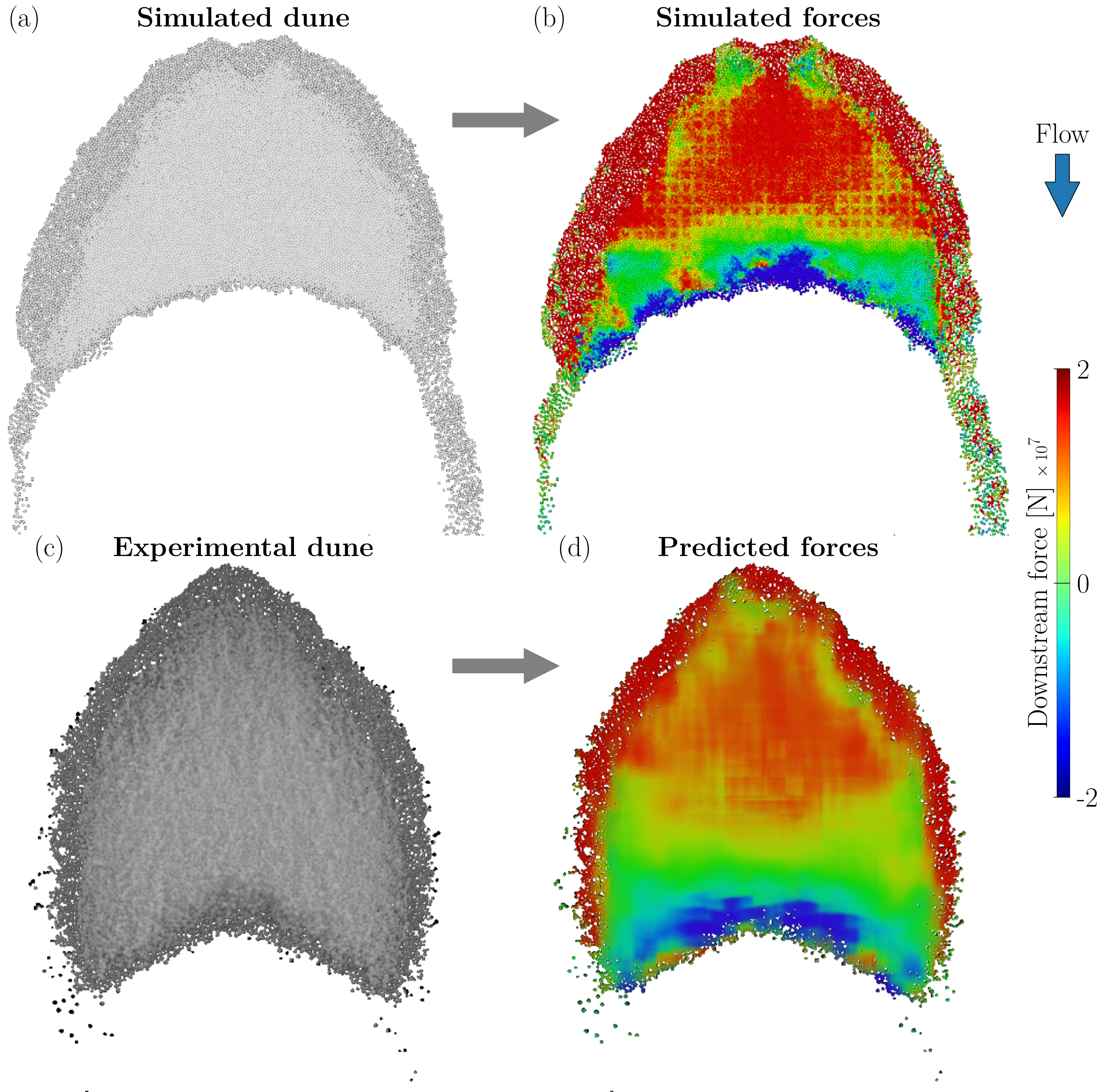}
	\caption{(a) Top view of subaqueous dune obtained from numerical simulation showing the dune topology. (b) Instantaneous forces over the barchan, computed by numerical simulation. (c) Top view of subaqueous dune obtained from experiment. (d) Instantaneous forces over the barchan, estimated by CNN.}
\label{fig:instantaneous}
\end{figure}

\section*{Estimating the resultant force on grains of real dunes}
\label{sec:results}

The backbone of this work consists of leveraging data from numerical simulations to increase the amount of information that can be extracted from experiments. Then, we aim to predict the forces on barchan dunes from experiments using a CNN model trained with simulation data. The force prediction is made only based on the information about the morphology of the dune found in the image.

The numerical computations of some of our previous work \cite{Alvarez5,Alvarez7,Lima2} are complemented with new ones, all using an Eulerian-Lagrangian approach where the fluid phase is computed using computational fluid dynamics (CFD) and the motion of individual grains is obtained by the discrete element method (DEM). Details of the numerical methods and setups are available in the Supplementary Material and in open repositories \cite{Supplemental_numericaldunes_pof,Supplemental_numericaldunes_pof2}. 
The experiments were conducted in a water tank where a single barchan dune is formed and entrained by the fluid flow for the same conditions as in the numerical simulations. In both the experiments and numerical simulations, the grains consisted of glass spheres with mean diameter $d$ $=$ 0.5 mm, the number of grains varied within 36,000 and 40,000, and the cross-sectional flow velocities varied between 0.294 and 0.364 m/s. For more details of the experimental setup, including a complete description and photographs, please see the Supplementary Material.
%In both the experiments and numerical simulations, the number of grains varies within 36,000 and 40,000. The grains consist of glass spheres with mean diameter $d$ $=$ 0.5 mm, and the cross-sectional flow velocities vary between 0.294 and 0.364 m/s. For more details of the experimental setup, including a complete description and photographs, please see the Supplementary Material. %Examples of top view images of numerically and experimentally generated barchans can be seen in Figs. \ref{fig:instantaneous}a and  \ref{fig:instantaneous}c, respectively. 

Figures \ref{fig:instantaneous}a and \ref{fig:instantaneous}b show a top view of a subaqueous dune and the distribution of the longitudinal component of the resultant force, respectively. These results were obtained by a numerical simulation and serve as input and output for training a CNN model that should be able to generalize results for experimental setups. Such outcomes can be observed in Figs. \ref{fig:instantaneous}c and \ref{fig:instantaneous}d, which show a top view of an experimental barchan and the prediction of the distribution of the longitudinal component of the resultant force acting on its grains. The background of the experimental dune is removed for visualization purposes. For all cases shown in this work, the flow direction is from top to bottom, as indicated by the blue arrow, and the hot colors indicate a positive force (in the flow direction) while the cold colors indicate a negative force. Hence, blue colors represent regions of flow recirculation, behind the dune.

To synthesize images of force distribution on dunes, such as that in Fig. \ref{fig:instantaneous}, we uniquely treat the image semantic segmentation as an image-to-image translation task that infers pixel-level labels of structures from the input images in a supervised way. This is done through a U-Net architecture \cite{Ronneberger_etal_2015}, a network developed to work with fewer training images and produce accurate image segmentation. 
A crucial step in the image-to-image translation task is to verify the model capacity, i.e., its ability to fit a wide variety of functions. Although a model with low capacity is incapable of solving complex tasks, it can overfit and produce poor generalization when its capacity is greater than necessary. Thus, we seek a model that parameterizes a function receiving the image of the dune, being able to predict the forces over its surface, while being able to generalize to cases of unseen dunes. We also \sout{want} aim to develop the capability to work with images obtained from different sources, for example, simulations and experiments. Naturally, we need to ensure that the generalizations are accurate enough in the sense that they obey or approximate the governing physics.

Determining the capacity of a deep learning model is especially difficult \cite{DeepLearningBook}. Here, we explore different datasets to assess how the predictions are impacted by adding or removing features from the training set (more details are provided in the Supplementary Material).
%
%% O texto abaixo foi enviado para o SM
%
%We define {\em domain} as the combination of an input space $\mathcal{X}$, an output space $\mathcal{Y}$ and their associated joint probability distributions. Since the flow and the corresponding dune morphology are influenced by the effects of Reynolds number and number of grains, we consider a collection of images from numerical simulations or experiments as forming a domain. However, for the case of experimental images, we have no access to $\mathcal{Y}$ during training\footnote{When only the inputs of the target domain are available, we have the setting of {\em domain adaptation}, while the term {\em domain generalization} refers to the case when no target domain data is accessed during training. This makes the latter more challenging than the former.}.
%
Considering an input space $\mathcal{X}$ and an output space $\mathcal{Y}$, we build a collection of datasets $\mathcal{D}_{s}^{i} = \{ (\mathbf{x}_j, \mathbf{y}_j) \}_{j=1}^{n} \in (\mathcal{X} \times \mathcal{Y})^n \ | \ i = \{1, 2, 3, 4\}$ composed of $n$ images obtained solely via simulation. As there are no force measurements for experimental dunes (we have no access to $\mathcal{Y}$ in the experimental domain), one can only verify the capacity of the model in the numerical domain $\mathcal{D} = \mathcal{D}_{s}$. Here, we use the subscript $s$ to refer to the domain of {\em simulation} data, while the upper indices indicate the simulations included in the individual datasets as listed in Table \ref{table:simulations}. The domain of {\em experimental} images is labeled here as $\mathcal{D}_{e}$.
\newcommand{\circleBlue}[1][blue,fill=blue]{\scalerel*{\tikz \draw[#1] (0,0) circle (4pt);}{\circ}}
\newcommand{\circleCyan}[1][cyan,fill=cyan]{\scalerel*{\tikz \draw[#1] (0,0) circle (4pt);}{\circ}}
\newcommand{\circleTeal}[1][teal,fill=teal]{\scalerel*{\tikz \draw[#1] (0,0) circle (4pt);}{\circ}}
\newcommand{\circleGray}[1][gray,fill=gray]{\scalerel*{\tikz \draw[#1] (0,0) circle (4pt);}{\circ}}
\newcommand{\circleBlack}[1][black,fill=black]{\scalerel*{\tikz \draw[#1] (0,0) circle (4pt);}{\circ}}
\definecolor{whitesmoke}{rgb}{0.96, 0.96, 0.96}
\begin{table*}[!htb]
\centering
% \begin{footnotesize}
\begin{tabular}{ c c c c | c c c c } 
    \hline
    \multicolumn{4}{c | }{Simulation parameters} & \multicolumn{4}{c}{Datasets} \\
    \hline
    \# & Reynolds & No. particles & No. images & $\mathcal{D}_{s}^{1}$ & $\mathcal{D}_{s}^{2}$ & $\mathcal{D}_{s}^{3}$ & $\mathcal{D}_{s}^{4}$ \\
    \hline
    \rowcolor{whitesmoke}
    1  & $1.47 \times 10^4$  & 40,000 & 1340 & & \circleCyan & \circleTeal & \circleGray \\
    \rowcolor{whitesmoke}
    2  & $1.57 \times 10^4$  & 40,000 & 3820 & \circleBlue & \circleCyan & \circleTeal & \circleGray \\
    \rowcolor{whitesmoke}  
    3  & $1.82 \times 10^4$  & 40,000 & 1500 & & & \circleTeal & \circleGray \\
    
    4  & $1.47 \times 10^4$  & 36,000 & 3730 & \circleBlue & \circleCyan & \circleTeal & \circleGray \\
    5  & $1.57 \times 10^4$  & 36,000 & 3838 & & \circleCyan & \circleTeal & \circleGray \\
    6  & $1.82 \times 10^4$  & 36,000 & 1760 & \circleBlue & & & \circleGray \\
    
    \hline
\end{tabular}
% \end{footnotesize}
\caption{Databases of high-fidelity simulations of fully-developed barchan dunes employed in this work and cardinality of the datasets. The colored circles identify the simulations that make up each dataset $\mathcal{D}_{s}^{i}$.}
\label{table:simulations}
\end{table*}

%
%% O texto abaixo foi enviado para o SM
%
%Naively training a model on an aggregate set of numerically simulated data pulled from the source domains can cause the model to learn domain-specific information that performs sub-optimally in experimental images, which will be our target domain. Variations in contrast, zoom, camera position, noise, intensity and direction of the light falling on the dune are frequent in experiments and configure a domain whose probability is very difficult to be faithfully represented in a simulation image. This is particularly important as perceptually insignificant changes in low-level statistics (natural scene statistics) can significant degrade the performance of the trained model \cite{1st ref in FDA}. %Note that we are not talking about the simulation being faithful to the physics of dune morphology, but rather the low-level statistics, which are informative about the structural complexity of a scene, being the same. 

In order to reduce the covariate shift\footnote{Covariate shift occurs when the input distribution changes between the training environment and the real one, negatively affecting the performance of the model.} caused by low-level statistics (natural scene statistics), we use data augmentation to force the model to learn possible photometric changes that can be found in an experimental image. These changes include rotation, translation and zoom. % (see Sec. \ref{sec:methods} for details).
Combined with this, we also reduce the dimensionality of the raw input to limit the number of relevant interacting factors that have to be learned, such as variations in contrast, noise, and intensity besides the direction of the light falling on the dune. This is justified by the fact that generalization is mostly achieved by a form of local interpolation among neighboring training examples \cite{2013_representation_learning}. Hence, we pre-treat the input images by applying a binarization to the grains through a threshold value and retrain the model with these binary inputs. %A sample of the final result is shown in Fig. \ref{fig:instantaneous_experiment}. The raw image on the left of the figure is binarized before entering the network to produce the output shown on the right.
To address the image imbalance in our simulation datasets (as shown in Table \ref{table:simulations}, the number of images per simulation varies), we apply an oversampling technique to ensure a more uniform distribution of images for the various flow setups.

\subsection*{Generalization to dunes never seen by the CNN model} \label{sec:generalize-dune}

\begin{figure}[!htb]
    \centering
    \begin{overpic}[width=0.99\textwidth]{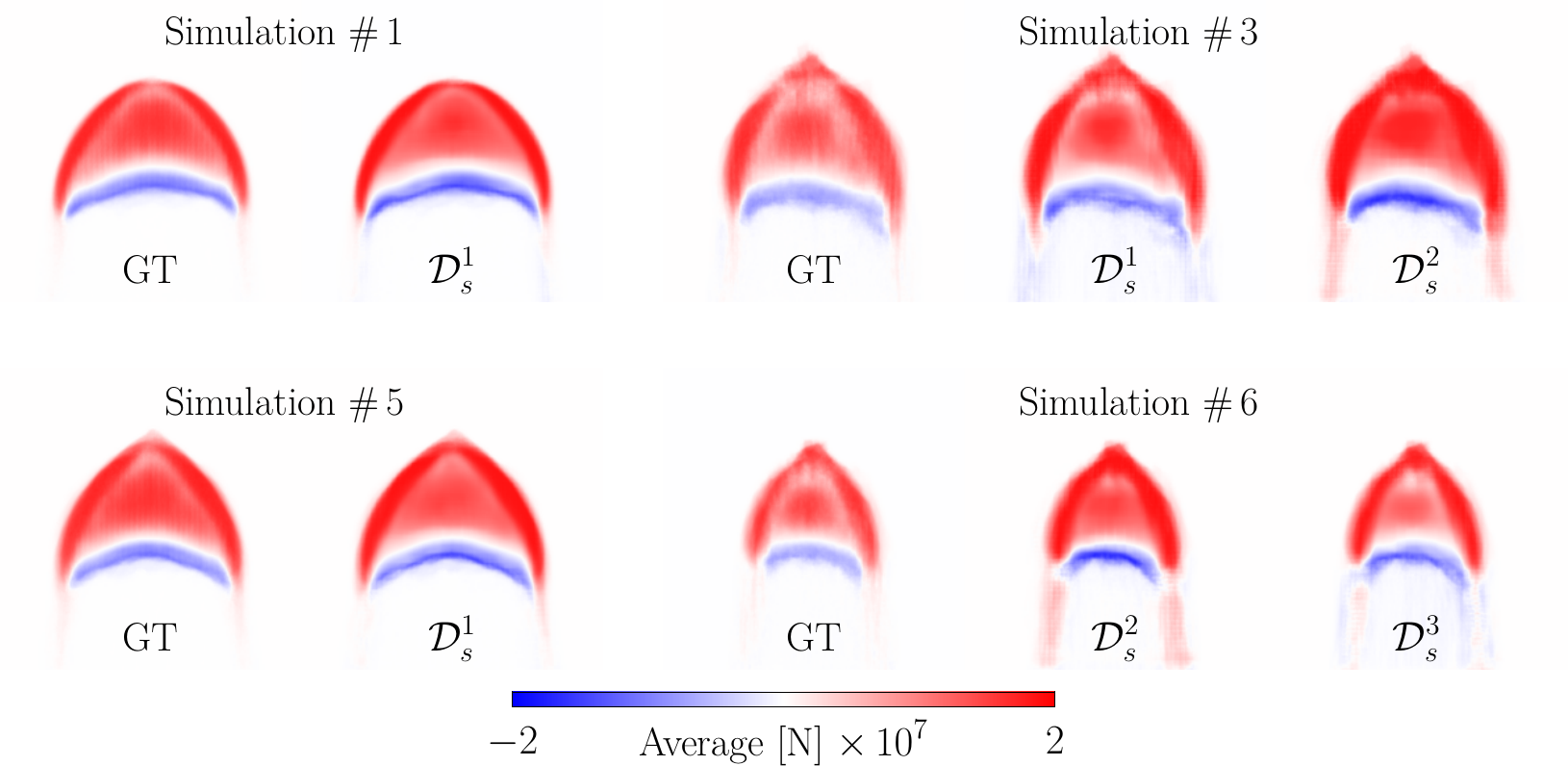}
        \put(2, 47){\textbf{\textcolor{black}{(a)}}}
    \end{overpic}
    \\
    \vspace{0.75cm}
    \begin{overpic}[width=0.99\textwidth]{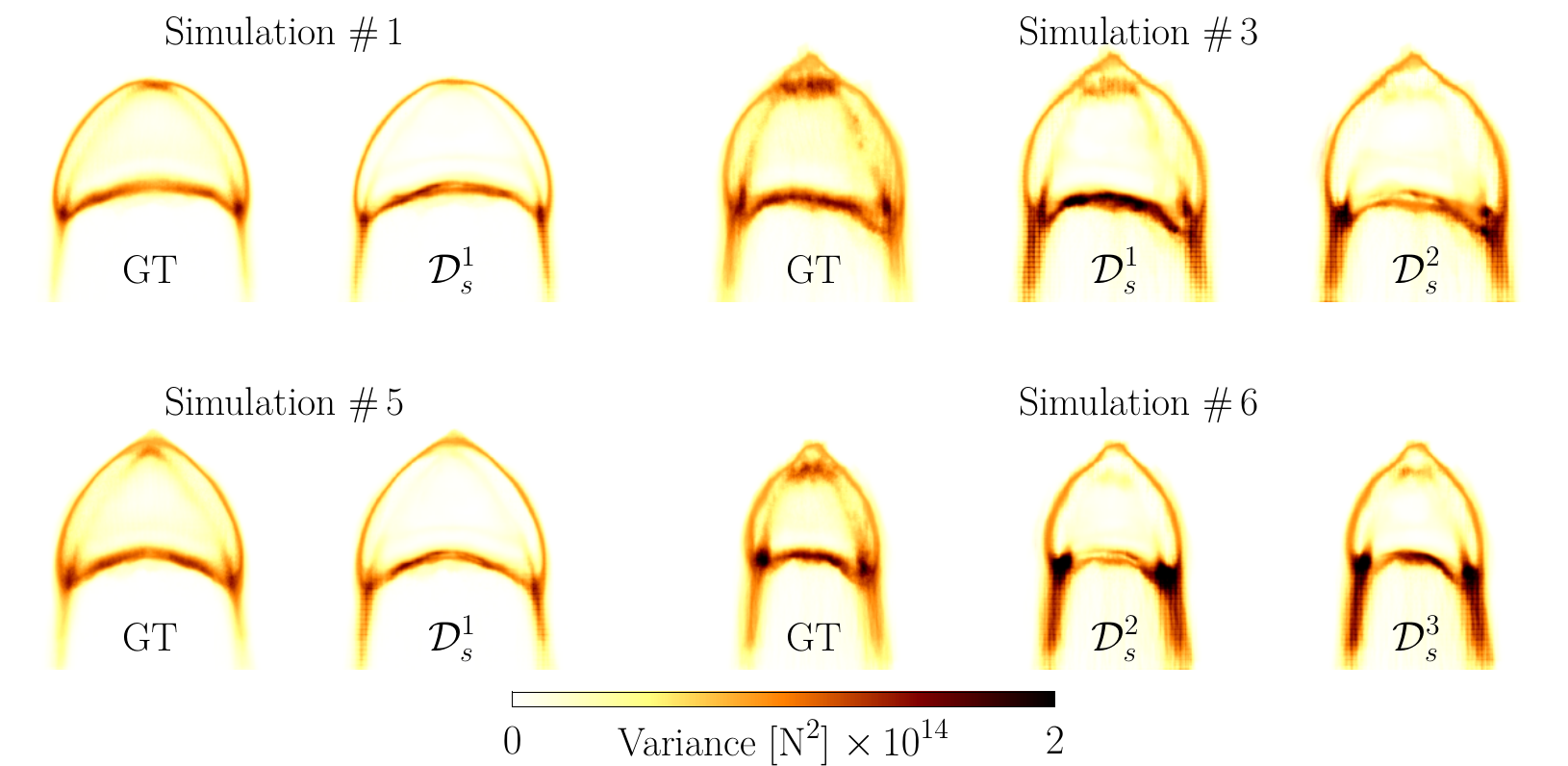}
        \put(2, 47){\textbf{\textcolor{black}{(b)}}}
    \end{overpic}
    \caption{Comparison of (a) time-averaged forces and (b) force variances on dunes from numerical simulations \#1, 3, 5 and 6. Results are shown for the ground truth (GT) and different CNN models trained with individual $\mathcal{D}_{s}^{i}$ datasets (see Table \ref{table:simulations}).}
    \label{fig:numerical}
\end{figure}

This section presents a comparison between results obtained by numerical simulations (ground truth) and the CNN models trained with different datasets $\mathcal{D}_{s}^{i}$ from Table \ref{table:simulations}. Figure \ref{fig:numerical} shows the time-averaged forces and their variances over numerical dunes extracted from simulations \#1, 3, 5, and 6. In order to minimize the effects of interval shift, we restrict the maximum and minimum levels of the force scale to constant values for the entire dataset. Here we use the range of $[-2, 2] \times 10^{-7}$ N, as shown in the force colorbar of Fig. \ref{fig:numerical}. To minimize posterior (concept) shift\footnote{Posterior shift happens when the distribution of outcomes changes, while the distribution of the input features remains the same.}, in turn, it is important to avoid color saturation in the images. We evaluate how the models trained with different datasets $\mathcal{D}_{s}^{i}$ perform in predicting dunes never seen. 

The dune shapes and sizes, as well as their shear forces depend on parameters such as the Reynolds number and the number of particles. From Table \ref{table:simulations}, one can infer that variations in these parameters lead to dunes with different morphologies. For instance, dunes \#3 and 6 have the same Reynolds number, but a different number of particles. As shown in Fig. \ref{fig:numerical}, the former is larger and has a sharper leading edge compared to the latter. On the other hand, dunes \#3 and 1 have the same number of particles, but a different Reynolds number. Such parameter change causes further considerable variations between the dune shapes. 

As we cannot explicitly determine {\em a priori} which set of features and variations will be relevant to determine the forces on the dune, it is convenient to express as many priors\footnote{A prior means the initial thought in terms of probability distribution over the parameters impacting the physical problem, for example, aspects related to the dune morphology.} 
about the dune morphodynamics as possible. 
Fortunately, the potential nuisance variability is usually already known in dune morphodynamics problems and can be dealt with early on, without the need to learn it through complex adversarial training \cite{domain_adversarial_nets2016, UDA-Review}. In fact, as observed in Fig. \ref{fig:numerical}, we know that the flow and corresponding dune response are influenced by the effects of Reynolds number and number of grains. Moreover, any color changes caused by the intensity of shear forces, including changes in the topology of the dune, are known nuisance factors and should be learned at the outset. This is especially important since the continuous interpretation of an image changes according to these factors, and networks do not transfer well across different feature and label distributions \cite{2019_low_level_transfer, survey_transfer_learning2010, UDA-Invariant-representation-2019}. 
Thus, including these sources of variability in the training set allows the model to extract and organize the discriminative information from the data that improves generalization, as demonstrated by the results.

Considering the parameters at hand, one can see from Fig. \ref{fig:numerical} that the mean forces and variances on dunes from simulations \#1 and 5 are well predicted by the CNN model trained with dataset $\mathcal{D}_{s}^{1}$. This model is informed with dunes containing the same number of grains and Reynolds numbers as the simulations, but for different combinations of such parameters. The same dataset is also able to reasonably estimate the forces over simulation \#3, while dataset $\mathcal{D}_{s}^{2}$ overpredicts the results. It should be pointed out that the latter dataset was never informed with a simulation at the same Reynolds number as simulation \#3, differently from the former one, which has results at the same Reynolds number but a different number of grains. This confirms the importance of the Reynolds number for an accurate force prediction. Similarly, for having a case with the same Reynolds number, the model trained with $\mathcal{D}_{s}^{3}$ provides a better estimate of the forces for simulation \#6 than $\mathcal{D}_{s}^{2}$. Finally, through training with high-fidelity numerical simulations at different conditions, the CNN is capable of interpolating results from experimental images generated from unseen flow parameters.
% Finally, the CNN can be applied to images from experiments carried out at similar conditions as the numerical simulations, as shown in Figs. \ref{fig:instantaneous}c and \ref{fig:instantaneous}d using the model $\mathcal{D}_{s}^{1}$. 
This opens the possibility of using the same technique for estimating the resultant force over dunes (and also other objects) imaged with remote sensing.

\section*{Conclusions}

% The proposed method represents a new step toward measurements of inaccessible quantities, for instance the forces on individual grains of sand dunes, opening new possibilities for remote sensing on Earth, Mars and other celestial bodies. The outcome of this research can have considerable impact on agriculture\cite{Baas}, housing \cite{WoodTV}, forest conservation \cite{Baas}, and planetary exploration, to cite but a few examples.

In this study, we have introduced a groundbreaking method for measuring the resultant forces acting on individual grains of sand within barchan dunes by leveraging deep learning techniques, particularly convolutional neural networks (CNNs). By combining subaqueous experimental data with high-resolution numerical simulations, we trained a CNN to predict the force distribution across the surface of barchan dunes based solely on their morphological features. This method marks a significant advancement in the field, offering a novel approach to measure quantities that were previously inaccessible at the grain scale. Ultimately, this research lays the foundation for a new era of granular mechanics measurement, where machine learning provides powerful insights into the dynamics of complex natural systems.

Through our approach, we demonstrated that the CNN could accurately predict the forces on dunes in both experimental and simulation settings, even for dunes with morphologies not included in the training set. This capacity to generalize to unseen dune configurations is crucial, as it allows the method to be applied across a broad range of landscapes, each subject to different flow conditions.
% In this sense, we provide evidence supported by theoretical foundation that the explanatory factors of variations behind data must be included in the trailing dataset for the model to generalize well.
We also highlighted the importance of learning sources of variability that enable the model to generalize effectively. By aligning the marginal label distribution across domains, we ensure that the model can adapt to diverse scenarios—a key condition for effective generalization when learning domain-invariant representations.

By extending the applicability of this technique to remote sensing data, our research has the potential for significant impact across various fields, including agriculture\cite{Baas}, housing\cite{WoodTV}, forest conservation\cite{Baas}, and planetary exploration. This opens up new possibilities for monitoring and understanding environmental and geological processes, both on Earth and beyond, and provides a powerful tool for addressing real-world challenges related to landscape dynamics and resource management.

% The flexibility and scalability of this approach pave the way for further advancements in the analysis of planetary surface interactions, contributing to a more comprehensive understanding of environmental and geological processes across different planetary bodies.

\bibliography{references}

\bibliographystyle{Science}

\section*{Acknowledgments}
The authors are grateful to FAPESP (Grants 2013/08293-7, 2018/14981-7, 2019/17874-0, 2021/06448-0, 2022/01758-3, 2022/09196-4). WRW and EMF are grateful to CNPq (Grants 308017/2021-8, 405512/2022-8, 444329/2024-2) for the financial support provided. 

% \ca{The Editor and reviewers will be acknowledged here after review.} 
% \ef{TO PLACE ANY OTHER ACKNOWLEDGMENT HERE.}

%Here you should list the contents of your Supplementary Materials -- below is an example. 
%You should include a list of Supplementary figures, Tables, and any references that appear only in the SM. 
%Note that the reference numbering continues from the main text to the SM.
% In the example below, Refs. 4-10 were cited only in the SM.  

\section*{Data and materials availability}

All data generated and used in this research, including the weights and biases of the trained model, have been made publicly available on Zenodo \cite{zenodo_dataset}. Additionally, the code used for processing the data, including the neural network model, is accessible through the GitHub repository of the Laboratory of Aeronautical Sciences. The repository can be found at \href{https://github.com/las-unicamp/measuring_forces_on_sand_dunes}{https://github.com/las-unicamp/measuring\_forces\_on\_sand\_dunes}.
Numerical results of the simulations, as well as the scripts used to post-process the numerical outputs are available on Mendeley Data \cite{Supplemental3} at \href{https://data.mendeley.com/datasets/zhnngnvvz8}{https://data.mendeley.com/datasets/zhnngnvvz8}. Setups for carrying out similar simulations using CFDEM are available on Mendeley Data \cite{Supplemental2} at \href{https://data.mendeley.com/datasets/ypkgwjfr4r}{https://data.mendeley.com/datasets/ypkgwjfr4r}

% https://github.com/las-unicamp/measuring_forces_on_sand_dunes
% Zenodo: 10.5281/zenodo.14262069

\clearpage
\section*{Supplementary materials}

The PDF file includes:
Materials and Methods\\
Supplementary Text\\
Figs. S1 to S10\\
Tables S1 to S4\\
References\\

\textbf{Other Supplementary Material for this manuscript includes the following:}\\
Movies S1 and S2\\

\subsection*{Materials and Methods}

\subsubsection*{Numerical simulations}

We carried out Euler-Lagrange simulations through CFD-DEM (computational fluid dynamics - discrete element method) computations. For that, we made use of the open-source code \mbox{CFDEM} (www.cfdem.com) \cite{Goniva}, which couples the CFD (computational fluid dynamics) open-source code OpenFOAM with the DEM (discrete element method) open-source code LIGGGHTS \cite{Kloss, Berger}. DEM is the Lagrangian part, solving, thus, the linear (Eq. \ref{Fp}) and angular (Eq. \ref{Tp}) momentum equations for each solid particle,

\begin{equation}
m_{p}\frac{d\vec{u}_{p}}{dt}= \vec{F}_{p}\,\, ,
\label{Fp}
\end{equation}

\begin{equation}
I_{p}\frac{d\vec{\omega}_{p}}{dt}=\vec{T}_{c}\,\, ,
\label{Tp}
\end{equation}

\noindent where, for each grain, $m_{p}$ is the mass, $\vec{u}_{p}$ is the velocity, $I_{p}$ is the moment of inertia, $\vec{\omega}_{p}$ is the angular velocity, $\vec{T}_{c}$ is the resultant of contact torques between solids, and $\vec{F}_{p}$ is the resultant force, given by Eq. \ref{Fp2}, 

\begin{equation}
\vec{F}_{p}= \vec{F}_{fp} + \vec{F}_{c} + m_{p}\vec{g}\,\, ,
\label{Fp2}
\end{equation}

\noindent where $\vec{F}_{fp}$ is the resultant of fluid forces acting on each grain, $\vec{F}_{c}$ is the resultant of contact forces between solids, and $\vec{g}$ is the acceleration of gravity. For $\vec{F}_{c}$, we considered Hertzian contacts in the normal and tangential directions, and we make reference to Lima et al. \cite{Lima2} for a detailed description. The resultant of fluid forces $\vec{F}_{fp}$ was computed as in Eq. \ref{Ffp_sim}, 

\begin{equation}
	\vec{F}_{fp} = \vec{F}_{d} + \vec{F}_{press} + \vec{F}_{\tau} + \vec{F}_{am} \,\, ,
	\label{Ffp_sim}
\end{equation}

\noindent where $\vec{F}_{d}$ is the drag force caused by the fluid, $\vec{F}_{press}$ is the force due to the pressure gradient, $\vec{F}_{\tau}$ is the force due to the gradient of the deviatoric stress tensor, and $\vec{F}_{am}$ is the added-mass force. In Eq. \ref{Ffp_sim}, we neglected the Basset, Saffman and Magnus forces since they are considered of low importance in CFD-DEM simulations \cite{Zhou}.

The CFD is the Eulerian part, solving the mass and momentum equations for the fluid. We used an unresolved approach, where the equations of motion are phase-averaged (volume basis) while assuring mass conservation, and we considered the equations of \textit{Set II} described in Zhou et al. \cite{Zhou}. Therefore, the mass and momentum equations are given by Eqs. \ref{mass_fluid} and \ref{momentum_fluid}, respectively,

\begin{equation}
	\frac{\partial \left( \alpha_{f} \rho_{f} \right)}{\partial t} + \nabla \cdot \left ( \alpha_{f} \rho_{f} \vec{u}_{f} \right ) = 0 \,\, ,
	\label{mass_fluid}
\end{equation}

\begin{equation}
	\frac{\partial \left ( \alpha_{f} \rho_{f} \vec{u}_{f} \right ) }{\partial t} + \nabla \cdot \left ( \alpha_{f} \rho_{f} \vec{u}_{f} \vec{u}_{f} \right ) = -\alpha_{f} \nabla P - \vec{f}_{exch} + \alpha_{f} \nabla \cdot  \vec{\vec{\tau}}_{f}  + \alpha_{f} \rho_{f} \vec{g} \,\, ,
	\label{momentum_fluid}
\end{equation}

\noindent where $\alpha_{f}$ is the volume fraction of the fluid, $\vec{u}_{f}$ is the fluid velocity, $\rho_{f}$ is the fluid density, and $\vec{f}_{exch}$ is the phase-averaged forces per unit of volume acting on solid particles,

\begin{equation}
	\vec{f}_{exch} = \frac{1}{\Delta V}\sum_{i}^{n_{p}} \left( \vec{F}_{d} +  \vec{F}_{am} \right) \, .
	\label{forces_exchange}
\end{equation}

\noindent where $n_p$ is the number of particles in the considered cell of volume $\Delta V$. In Eq. \ref{momentum_fluid}, $\vec{F}_{\tau}$ and $\vec{F}_{press}$ are separated from the remaining forces and appear explicitly in the equation, and more details on the computation of $\vec{f}_{exch}$ can be found in Lima et al. \cite{Lima2}. In OpenFOAM, Eqs. \ref{mass_fluid} and \ref{momentum_fluid} are divided by the fluid density, and the pressure term is computed separately. Therefore, the resulting term $\nabla \cdot  \vec{\vec{\tau}}_{f} / \rho$ becomes

\begin{equation}
	\alpha_{f} \nabla \cdot  \vec{\vec{\tau}}_{f} / \rho_f = \nabla \cdot \left( \alpha_f(\nu + \nu_t)\nabla\vec{u}_{f} \right) + \nabla \cdot \left( \alpha_f(\nu + \nu_t) \nabla ( \vec{u}_{f} )^T - \frac{2}{3} (\nabla \cdot \vec{u}_{f}) \vec{\vec{I}}\right) \,\,,
	\label{eq:stressTensor}
\end{equation}

\noindent where $T$ stands for transposed, $\vec{\vec{I}}$ is the identity tensor, $\nu$ is the kinematic viscosity of the fluid, and $\nu_t$ is the sub-mesh turbulent viscosity.

The numerical domain consisted of a rectangular-cross sectional channel with 0.3 m $\times$ 0.05 m $\times$ 0.16 m in the streamwise, vertical, and spanwise directions, respectively. For the fluid (CFD part), we made use of large-eddy simulations (LES) with the wall-adapting local eddy-viscosity (WALE) approach. Impermeability and no-slip boundary conditions were applied at the top and bottom walls, with periodic conditions in the longitudinal and transverse directions. For the particles, the boundary conditions consisted of solid walls at the top and bottom walls, free exit at the outlet, and no mass entry at the inlet. Prior to the CFD-DEM computations, we carried out LES simulations of a single-phase water flow. After reaching a fully-developed turbulent flow, the LES simulations were stopped and the results stored to be used as initial condition for the fluid. For the grains, we used glass spheres with mean diameter $d$ = 0.5 mm, which we let settle by free fall on the bottom wall of the channel (filled with still water), forming one conical pile. The pile, consisting of 36,000 or 40,000 particles (depending on the simulation) and centered 0.05 m from the domain inlet, had a radius of 25 mm. The CFD-DEM simulations then began by imposing the single-phase water flow previously stored. By varying the cross-sectional mean velocity of the water, $U$, we simulated three different flow conditions characterized by the channel Reynolds number $Re = U 2\delta / \nu$, where $2\delta$ $=$ 0.05 m corresponds to the channel height. The values of the Reynolds number and the number of particles for each simulation are provided in Table \ref{table:simulations}. This numerical approach has been extensively validated through comparisons with experimental results in Refs. \cite{Alvarez5,Alvarez7}. Numerical setups, outputs, and scripts for post-processing the outputs are available in open repositories \cite{Supplemental3, Supplemental2}.

\subsubsection*{Experiments}

\begin{figure}[ht]
	\centering
	\includegraphics[scale=0.75]{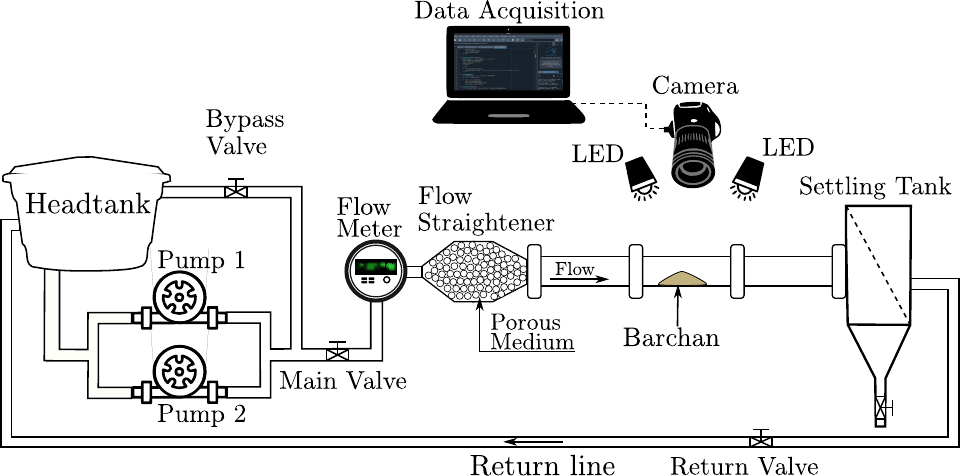}
	\caption{Sketch of the experimental setup. The experiment begins with a head tank that serves as the primary source of water or fluid, ensuring a stable and consistent flow. Two pumps circulate the fluid continuously, while a bypass valve regulates excess flow and a main valve directs the fluid through the system. Before reaching the experimental section, the fluid passes through a flow straightener, a porous medium used to homogenize the flow. A digital flow meter precisely measures the flow rate, allowing for accurate control and monitoring throughout the experiment. Any sediment or particulates in the fluid are collected in a settling tank, ensuring that the fluid returned to the system is clean and suitable for recirculation. The return line and return valve guide the fluid back to the pumps, maintaining the system's closed-loop operation. A data acquisition system, consisting of a computer connected to a high-resolution camera, collects and processes experimental data. The camera, positioned for a top view, captures detailed images of the experiment, with LED lights providing uniform illumination to enhance visualization.}
    \label{fig_layout}
\end{figure}

The experimental setup consisted of a water reservoir, two centrifugal pumps, a flow straightener, a 5-m-long closed-conduit channel of rectangular cross section, a settling tank, and a return line. The channel was 160 mm wide and 50 mm high, the last 2 m of which corresponded to the 1-m-long test section followed by a 1-m-long section discharging in the settling tank. Therefore, the entrance length was of approximately 40 hydraulic diameters, assuring a fully-developed water flow at the test section. The controlled grains were allowed to settle by free fall on the bottom wall of the test section (already filled with still water), forming a conical heap that was later deformed into a barchan shape by imposing a turbulent water flow. Figure \ref{fig_layout} shows a layout of the experimental setup.

\begin{figure}[ht]
	\centering
	\includegraphics[scale=1.4]{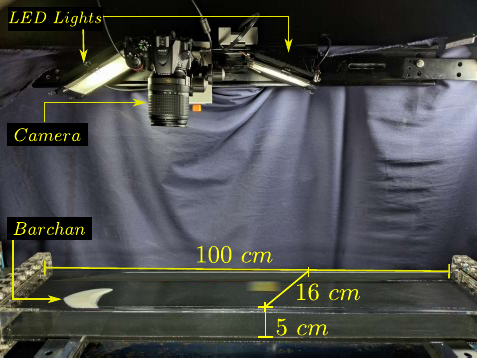}
	\caption{Photograph of the experimental setup. On the bottom, the photograph shows the test section with a barchan dune inside. On the top, the photograph shows the LED lights and the camera.}
    \label{fig_setup_exp}
\end{figure}
	
A high-speed camera was mounted on a traveling system, on the top of the test section, capturing top-view images of the barchan as it moved through the channel. The camera is of complementary metal-oxide-semiconductor (CMOS) type, with maximum resolution of 2560 px $\times$ 1600 px at 1000 Hz, and was assembled with a 60 mm-focal-distance lens of F2.8 maximum aperture. The region of interest (ROI) was fixed to 2560 px $\times$ 1304 px, and the acquiring frequency to 70 Hz. The selected region provided a good resolution for the the grains, while keeping the entire dune in the field of view of the camera, and corresponded to a spatial resolution of $20$ px/mm. To prevent beating between the camera and lighting, we used LED lamps connected to a constant current source. Figure \ref{fig_setup_exp} shows a photograph of the test section of the experimental setup.

\begin{table}[!ht]
	\centering
	\begin{tabular}{|l|c|c|c|c|c|c|}
		\hline
		Case   & $Re$    & $U$ (m/s) & $u_*$ (m/s) & $N$ & $m_0$ (g) \\ \hline
		I   & 14,700 & 0.294  & 0.0168 & 40,000    & 7.0        \\ \hline
		II  & 15,700 & 0.314  & 0.0176 & 40,000    & 7.0        \\ \hline
		III & 18,200 & 0.364  & 0.0202 & 40,000    & 7.0        \\ \hline
		IV  & 14,700 & 0.294  & 0.0168 & 36,000    & 6.0        \\ \hline
		V   & 15,700 & 0.314  & 0.0176 & 36,000    & 6.0        \\ \hline
	\end{tabular}
	 \caption{\label{tab:exp_scketch_} Test conditions of the experiments: Case number, Reynolds number $Re$, cross-sectional mean velocity $U$, shear velocity $u_*$, number of solid particles $N$, and initial mass of the dune $m_0$.}
\end{table}

\begin{figure}[ht]
	\centering
	\includegraphics[scale=1.0]{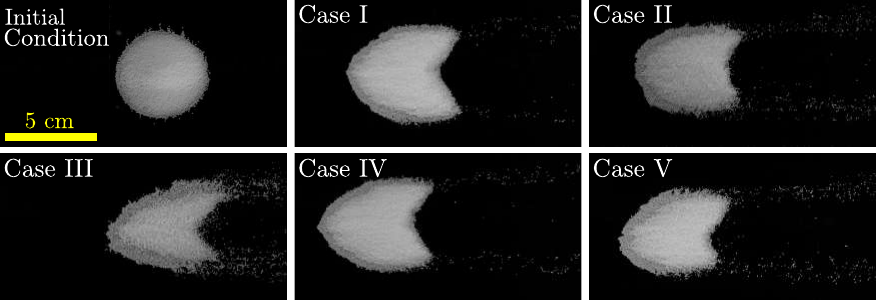}
	\caption{\label{fig:case_i_to_v_frame_500} Snapshots showing top views of bedforms for the typical initial condition and developed dunes for the cases I to V, as listed in Table \ref{tab:exp_scketch_}. The corresponding cases are listed on the top left of each panel.}
    \label{fig_snapshots_exp}
\end{figure}

Five tests were performed using tap water at temperatures within 24 and 26 $^\circ C$ and round glass spheres ($\rho_g$ $=$ 2500 kg/m$^3$) with 0.40 mm $\leq$ $d$ $\leq$ 0.60 mm, where $\rho_p$ and $d$ are, respectively, the density and diameter of the glass spheres. The cross-sectional mean velocities U varied within 0.294 and 0.364 m/s, corresponding to 1.47 $\times$ 10$^4$ $\leq$ $Re$ $\leq$ 1.82 $\times$ 10$^4$, as shown in Tab.~\ref{tab:exp_scketch_}. The shear velocities $u_*$ on the channel walls were computed from velocity profiles acquired by a two-dimensional particle image velocimetry device, and were found to follow the Blasius correlation \cite{schlichting2000boundary}. Figure \ref{fig_snapshots_exp} shows top views of the the initial bedform as well as the developed barchans for cases I to V.

\subsubsection*{CNN model}

As mentioned in the text, to synthesize images of force distribution on dunes, such as that in Fig. \ref{fig:instantaneous}, we uniquely treat the image semantic segmentation as an image-to-image translation task that infers pixel-level labels of structures from the input images in a supervised way. This is done through a U-Net architecture \cite{Ronneberger_etal_2015}, a network developed to work with fewer training images and produce accurate image segmentation. The U-Net combines a pixel-wise Softmax over the final feature map with the cross entropy loss function. However, the Softmax is removed in the present framework and the mean squared error (MSE) loss function is employed instead. 
\begin{figure}[!htb]
    \centering
    \includegraphics[width=0.99\textwidth]{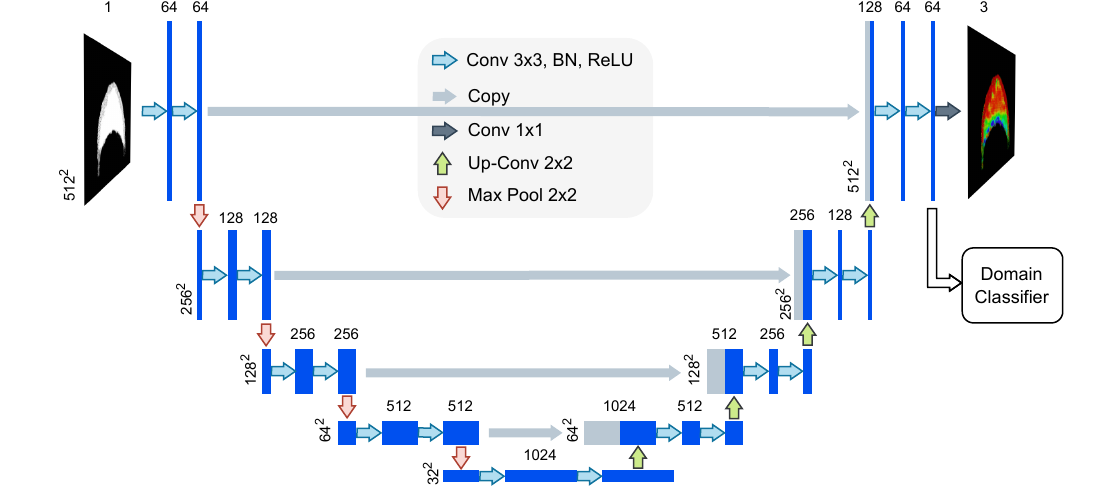}
    \caption{Diagram of the U-Net architecture coupled with a domain adversarial discriminative module.}
    \label{fig:unet}
\end{figure}

Naively training a model on an aggregate set of numerically simulated data pulled from the source domains\footnote{We define {\em domain} as the combination of an input space $\mathcal{X}$, an output space $\mathcal{Y}$ and their associated joint probability distributions.}\footnote{The \em{source} domain refers to the domain where labeled data is available and used to train a model, while the \em{target} domain is the domain where the model needs to be applied, but labeled data may be scarce or unavailable.}
can cause the model to learn domain-specific information that performs sub-optimally in experimental images, which is our target domain. Variations in contrast, zoom, camera position, noise, intensity and direction of the light falling on the dune are frequent in experiments and configure a domain whose probability is very difficult to be faithfully represented in a simulation image. This is particularly important as perceptually insignificant changes in low-level statistics (natural scene statistics) can significant degrade the performance of the trained model \cite{2019_low_level_transfer}. Note that we are not talking about the simulation being faithful to the physics of dune morphology, but rather the low-level statistics, which are informative about the structural complexity of a scene, being the same.

There is the intuition that deep learning algorithms can extract features that disentangle the underlying factors of variations, helping to perform transfer across domains (see \cite{2011_DA_classification} for instance). This is justified by the fact that the intermediate, abstract concepts learned by the model are general enough to make sense across a wide range of domains, which translates into better transfer \cite{2014_Learning_and_transferring_midlevel, 2013_representation_learning}. However, erroneous predictions can be made by deep models even when subtle changes from the training domain occur \cite{2018_conditional_adversarial_DA, UDA-Invariant-representation-2019}. As deep representations can only reduce, but not remove the cross-domain distribution discrepancy \cite{2018_conditional_adversarial_DA}, adaptation modules are embedded in deep networks for distribution matching. The development of such adaptation modules represent an open problem in the literature of deep learning.

Denoting $X$ and $Y$ as the image features and target (dune response), respectively, domain adaptation mainly considers the covariate shift situation, trying to find invariant components $\mathcal{T}(X)$ that have similar $P(\mathcal{T}(X))$ on different domains by explicitly minimizing a distribution discrepancy measure, such as maximum mean discrepancy or correlation distances, or through adversarial training \cite{2014_domain_confusion, 2015_deep_adaptation_nets} or even self supervision \cite{2020_single_source_UDA}. However, because there are no labels in the target domain, the shared representation $\mathcal{T}$ cannot be learned by minimizing the distance between source $P^{S}(Y|\mathcal{T}(X))$ and target $P^{T}(Y|\mathcal{T}(X))$, and it is not clear if these conditional distributions will be similar. In other words, there is no guarantee that the learned marginally invariant representation has sufficient structural (semantic) information to be conditionally invariant. In fact, the assumption of marginal distribution invariance is not only hardly met, but also it has been shown to perform sub-optimally for classification problems \cite{2021_DA_with_invariant_representation_learning, UDA-Invariant-representation-2019} and has been considered insufficient \cite{2019_Supp_and_Inv_in_DA}.

Similar conclusions were also brought by exploiting the causal mechanism of the data-generating process \cite{2021_DA_with_invariant_representation_learning}. According to \cite{2018_DG_via_conditional_adversarial_net}, in a causal structure $A \rightarrow B$, the mechanism $P(B|A)$ is independent of the cause generating process $P(A)$ and so it remains stable as $P(A)$ changes. However, in computer vision the causal structure is often $B \rightarrow A$, which means that changes in $P(A)$ reflects on $P(B|A)$. That is, if the invariance of the conditional distribution $P(Y|X)$ is violated, the joint distribution $P(Y, \mathcal{T}(X))$ will not be invariant even if $P(\mathcal{T}(X))$ is invariant after learning.

The fundamental limitation of domain-invariant representation is the potential discrepancy between the marginal label distribution \cite{UDA-Invariant-representation-2019}. Recently, under the assumption that the label distribution is unchanged across domains, \cite{2021_Domain_density_transformations} proposed a way of learning domain-invariant representations that align both marginal -  $P(\mathcal{T}(X))$ - and conditional - $P(Y|\mathcal{T}(X))$ - distributions, while also showing that the invariance of the distribution of class labels across domains is a necessary and sufficient condition for the existence of domain-invariant representations.

Despite the recent theoretical advances mentioned above, most domain algorithms concern classification problems. But in our work, we seek finding the parameterized function $\Phi : \mathbb{R} \rightarrow \mathbb{R}^3, \ g(\mathbf{x}) \mapsto f(\mathbf{x}) \ | \ \mathbf{x} \in \mathbb{R}^n$ to map the scalar value $g(\mathbf{x}_i)$ in each spatial pixel coordinates $\mathbf{x}_i = (x_i, \ y_i)$ to its associated RGB color levels $f(\mathbf{x}_i)$ representing the forces (notice that here we are using $x$ and $y$ to denote the pixel coordinates and not the input and output of the network). Hence, this is a high-dimensional regression task for which, to the authors knowledge, there is no adaptation module readily available\footnote{We acknowledge the pioneering work of \cite{2022_MSE_for_visual_regression} that provides a solution to multi-dimensional imbalanced regression, however, the dimensionality of our problem greatly exceeds that considered by them.}. Even one-dimensional regression tasks still lack an effective approach as they are much more difficult than their classification counterpart. This difficulty arises from several facts, including the fact that there are no hard boundaries between targets and their distances have a meaning (i.e., targets have an ordinal data structure \cite{2020_Continuously_indexed_DA, 2021_UDA_in_ordinal_regression}); that of the imbalance of target values, which requires interpolation and extrapolation \cite{2021_Deving_into_DIR, 2022_MSE_for_visual_regression}; as well as the regression not being robust to feature scaling \cite{2021_Representation_subspace_distance_for_DAR}. While in both classification and regression tasks domains may have different probability densities (imbalanced domain regression), a disjoint interval (label) range may also occur in regression tasks.

In order to minimize covariate shift effects, we employed the gradient reversal algorithm proposed by \cite{2015_UDA_with_backprop} that jointly optimizes the underlying features as well as a discriminative domain classifier operating on these features through an adversarial loss. Hence, it treats domain invariance as a binary classification problem for the domain and this domain classifier encourages domain-invariant features to emerge in the course of optimization. We observed, however, that this approach was not sufficient to infer correct results in experimental images. Despite not knowing what the true forces acting on the experimental really are, we expect them to be similar to those from the simulations, in the sense that the convex peripheral region of the dune is red, due to strong shearing forces, and the concave part is predominantly blue due to the fluid recirculation zone. But, that is not what we saw in the output of the neural network (now shown for brevity). Therefore, we chose to apply discriminative feature extraction through binarization of the input images, directing the network's focus toward the structural features, such as the shape and contours of the dunes.

Determining the capacity of a deep learning model is especially difficult \cite{DeepLearningBook}, and we do not intend to provide generalization error bounds to quantify it here. This is an active area of research that is outside the scope of this work. Instead, we explore different datasets to assess how the predictions are impacted by adding or removing features from the training set. The idea is that the model can generalize well only if it is provided with an appropriate feature space such that it can capture enough of the complexity of interest and disentangle the underlying factors of variation. Hence, leveraging a large number of labeled training datasets from various domains becomes a promising solution. Since the flow and the corresponding dune morphology are influenced by the effects of Reynolds number and number of grains, we consider a collection of images from numerical simulations or experiments as forming a domain. However, for the case of experimental images, we have no access to $\mathcal{Y}$ during training\footnote{When only the inputs of the target domain are available, we have the setting of {\em domain adaptation}, while the term {\em domain generalization} refers to the case when no target domain data is accessed during training. This makes the latter more challenging than the former.}.

\end{document}